%% file: main.tex
\documentclass[sigconf]{acmart}

\usepackage{booktabs}
\usepackage{graphicx}
\usepackage{amsmath}

\begin{document}

\title{Bridge Evidence: Static Retrieval Utility Does Not Predict Causal
       Utility in Multi-Step Agentic Search}

\author{Debayan Mukhopadhyay}
\affiliation{%
    \institution{University of Calcutta}
    \city{Kolkata}
  \state{West Bengal}
    \country{India}
}
\email{debayan.mukherjee14@gmail.com}

\author{Utshab Kumar Ghosh}
\affiliation{%
  \institution{Missouri University of Science and Technology}
  \city{Rolla}
  \state{MO}
  \country{USA}
}
\orcid{0000-0003-3096-6909}
\email{u.ghosh@mst.edu}

\author{Shubham Chatterjee}
\affiliation{%
  \institution{Missouri University of Science and Technology}
  \city{Rolla}
  \state{MO}
  \country{USA}
}
\orcid{0000-0002-6729-1346}
\email{shubham.chatterjee@mst.edu}

\renewcommand{\shortauthors}{Utshab Kumar Ghosh, Debayan Mukhopadhyay and Shubham Chatterjee}

\input{sections/abstract}



\maketitle

\input{sections/introduction}
\input{sections/related_work}
\input{sections/methodology}
\input{sections/results}
\input{sections/conclusion}

\bibliographystyle{ACM-Reference-Format}
\bibliography{references}

\input{sections/appendix}

\end{document}

%% file: sections/abstract.tex
\begin{abstract}
Retrieval systems are trained and evaluated on a static idea of usefulness. Hand a document and a question to a reader model, see whether the answer improves, and score the document accordingly. Almost every standard metric reduces to some version of that question. The idea holds up when a document is read on its own. It starts to break when a language model works as a search agent, issuing several queries and reasoning across turns, because a document can matter for what it lets the agent do next rather than for what it says about the question in front of it.

We measure that gap directly instead of arguing about it. Using a ReAct style agent over HotpotQA, we replay a stratified sample of 1000 development questions and, for every document the agent actually read, delete it and re-run the rest of the trajectory from that point. Comparing the original run against its counterfactual gives a Counterfactual Trajectory Utility (CTU) score built from three deltas: the quality of the final answer, the retrieval quality of the query the agent issues next, and the number of turns it needs. Crossing CTU against Static RAG Utility (SRU) over 23{,}322 document observations, the two turn out to be close to statistically independent (Spearman $\rho = -0.026$). Roughly a third of the documents the agent reads are causally load bearing while looking useless to a static reader. We call these bridge documents. The pattern survives when we swap the reader based SRU axis for a BM25 and cross encoder proxy, which gives a bridge cell of 27.2\% with an evenly spread axis.

A second experiment pins down the mechanism. Using the Observable Entity Relevance (OER) measure from prior work on entity ranking, we find that entities whose presence discriminates relevant from non-relevant candidates appear in the agent's next query 4.02 times more often than entities found only in non-relevant documents (6.1\% against 1.5\%, $n = 227{,}139$). A bridge document earns its keep by handing the agent a discriminative entity that redirects the search. Taken together the results say that static relevance and causal usefulness are different quantities in agentic retrieval, and that optimizing the first does not deliver the second.
\end{abstract}

%% file: sections/introduction.tex
\section{Introduction}
\label{sec:introduction}

Ask a modern retrieval system what makes a document good and you get a clear answer. A document is good if handing it to a reader along with the question makes the answer better. That definition is doing an enormous amount of work. It sits underneath nDCG, MAP, and MRR~\citep{jarvelin2002cumulated}, underneath the training objective of most learned rankers~\citep{karpukhin2020dense,nogueira2019passage}, and underneath the way retrieval augmented generation is usually evaluated~\citep{lewis2020retrieval}. It also assumes something specific about how the document gets used, namely that a reader sees the question and the document together, with no history and no plans.

That assumption quietly stops being true once a language model is allowed to act. A ReAct style agent~\citep{yao2023react} reads a question, decides what it does not know, searches for it, reads what comes back, and searches again. Systems in this family now handle a large share of multi-hop question answering~\citep{trivedi2023interleaving,press2023measuring,asai2024selfrag}. In that setting the link between a document and the final answer runs through several turns of reasoning, and it stops being direct. A document can be the reason the agent succeeds without containing anything that resembles the answer. Another document can contain the answer outright and change nothing at all, because the agent already knew it.

Consider a question our agent actually faced. \emph{What political party was the 6th governor of Hawaii that passed bills to help with global warming?} At the first step the agent retrieves a passage saying that Linda Lingle served as the 6th Governor of Hawaii from 2002 to 2010. The passage says nothing about any political party and nothing about environmental legislation. Score it the standard way, by giving it to a fresh reader with the question, and it is worthless. The reader cannot answer from it. Yet that passage is the reason the agent succeeds, because it supplies the name Linda Lingle, which the agent puts into its next query, which retrieves the passage that does contain the party. Delete the first passage and the agent never learns whose party to look up. Its value lies entirely in what it unlocks.

We call documents of this kind \textbf{bridge evidence}. They bridge the gap between what the agent knows now and the next search worth running. Our claim is that they are common, that they are invisible to the signals retrieval systems currently optimize, and that this invisibility is systematic rather than accidental. A bridge document is defined by what it enables in the future, and every standard signal scores it on what it contains in the present.

\subsection{Measuring the gap instead of asserting it}

Arguments of this shape are easy to make and hard to believe without numbers, so we measure the gap. The tool is a counterfactual. If we want to know whether a document mattered causally, the cleanest thing to do is take it away and see what happens~\citep{pearl2009causality}.

We build a framework we call Counterfactual Trajectory Exploration. For every document the agent actually read at every step, we delete that one document from the ranked list, hand the agent the remaining evidence, and replay the rest of the trajectory from that point with everything else held fixed. The original run and the counterfactual run then differ in exactly one thing. Comparing them gives a Counterfactual Trajectory Utility (CTU) score, which combines three deltas: whether the final answer got worse, whether the query the agent issued next retrieved less gold evidence, and whether the agent needed more turns.

Placing CTU on one axis and Static RAG Utility (SRU) on the other turns a rhetorical claim into a contingency table. If the field's assumption is right, the two should agree and the table should be diagonal. It is not. Across 23{,}322 document observations from a stratified sample of 1000 HotpotQA development questions, the rank correlation between the two is $-0.026$. They are, to a good approximation, measuring unrelated things.

\subsection{Why bridge documents work}

Showing that a gap exists is one thing. Explaining it is another, and a reviewer is entitled to ask what a bridge document actually does. Our answer is that it introduces an entity. The agent picks that entity up and puts it in its next query, and the search changes direction.

That explanation is testable, and it connects to a separate line of work on entity relevance~\citep{ghosh2026entity}. That work distinguishes two notions. Conceptual Entity Relevance is the topical judgment a person or a language model makes about whether an entity is about a query. Observable Entity Relevance (OER) asks a sharper question: does the entity's presence in a candidate document actually separate relevant documents from non-relevant ones? The two agree only weakly, and OER is the one that helps retrieval. If our account of bridge evidence is right, then the entities that get carried into the next query should be the observably discriminative ones rather than whatever happens to be printed on the page. We test that prediction and it holds, by a factor of four.

\subsection{Contributions}

\begin{itemize}
  \item We formalize Counterfactual Trajectory Exploration for multi-step retrieval agents and give a concrete omission intervention that isolates the causal contribution of a single document while holding the rest of the trajectory fixed (Section~\ref{sec:method-cte}).
  \item We define Counterfactual Trajectory Utility, a composite of answer, next query, and effort deltas, and show that its natural threshold coincides exactly with the point of zero causal effect, which gives the high and low split a principled meaning rather than an arbitrary one (Section~\ref{sec:method-ctu}).
  \item We run the quadrant experiment over 23{,}322 document observations and find that static utility and causal utility are close to independent ($\rho = -0.026$), with 35.7\% of read documents landing in the bridge cell. The finding survives a robustness check that replaces the reader based axis with a BM25 and cross encoder proxy (Section~\ref{sec:results-quadrant}).
  \item We identify the mechanism. Entities with high Observable Entity Relevance propagate into the agent's next query 4.02 times more often than non-discriminative ones over 227{,}139 observations (Section~\ref{sec:results-propagation}).
  \item We report the limits of the evidence plainly, including a skew in the static utility axis that constrains how much the headline quadrant number can carry on its own (Section~\ref{sec:results-threats}).
\end{itemize}

%% file: sections/methodology.tex
\section{Methodology}
\label{sec:methodology}

\subsection{Problem definition}
\label{sec:method-problem}

Our starting claim is that causally essential evidence in multi-step agentic retrieval has a different shape than the evidence that existing proxy signals select. The proxies we have in mind are the familiar ones: the BM25 score, static reader utility, and whether the document contains the answer string. We argue the mismatch is not noise. It reflects a specific class of evidence.

\begin{quote}
\textbf{Bridge evidence.} A document that introduces an entity or a relation which appears in the agent's next query but was absent from every previous query, and whose removal degrades the trajectory, while the document itself scores low on existing static proxies.
\end{quote}

Bridge evidence is invisible to current signals for a structural reason rather than an accidental one. Its usefulness is defined by what it enables at the next step, and every static proxy scores a document on what it contains relative to the current query. In concrete terms we expect to find documents that score low on BM25, low on static reader utility, and low on answer containment, but whose deletion makes the agent fail or take noticeably longer.

\subsection{Trajectory representation}
\label{sec:method-trajectory}

We study a ReAct style research agent~\citep{yao2023react} that answers questions by searching Wikipedia. It reasons inside a thought block before each action, searches with an explicit query, and answers when it judges the evidence sufficient. It is capped at four search turns and instructed never to repeat a query. The full system prompt is reproduced in Appendix~\ref{app:prompts}.

The agent moves from state $S_t$ to state $S_{t+1}$ through the transition
\begin{equation}
  S_t \;\longrightarrow\; \big(\,(\text{thought} + \text{CW})_t,\; q_t,\; D_t\,\big) \;\longrightarrow\; S_{t+1},
  \label{eq:transition}
\end{equation}
where $S_t$ is the latent reasoning state, $(\text{thought} + \text{CW})_t$ is the explicit chain of thought together with the context window the agent has accumulated so far, $q_t$ is the sub-query the agent generates at step $t$, and $D_t$ is the evidence set the retriever returns for it. The next state is conditioned on all of these jointly. The agent's decision therefore depends not only on the individual documents $d_i \in D_t$ but also on the context window holding every fact and reasoning trace from earlier steps.

The retrieval backend returns the top 50 paragraphs for each query, of which the top 10 enter the agent's context. Query generation at the next step can be written as
\begin{equation}
  \big(\,(\text{thought} + \text{CW})_t,\; D_t[:10]\,\big) \;\longrightarrow\; q_{t+1}.
  \label{eq:querygen}
\end{equation}
Equation~\ref{eq:querygen} is the part of the loop this paper is really about. It is the point where a document that answers nothing can still change everything, by supplying the term that makes the next query work.

We do not intervene on the latent reasoning itself. It is not observable and not cleanly manipulable. We intervene on the evidence conditioning variable $D_t$, which is both observable and exactly the thing a retrieval system controls.

\subsection{Experimental setup}
\label{sec:method-setup}

\paragraph{Agent and retrieval stack.}
The agent backbone is Qwen2.5-7B-Instruct~\citep{qwen2025technical}. First stage retrieval is BM25~\citep{robertson2009probabilistic} over a paragraph level Wikipedia index, and the top 50 candidates are reranked by a cross encoder before the top 10 are shown. To keep the counterfactual comparison meaningful we hold decoding deterministic, with temperature set to $0.0$ and sampling disabled, and we run every trajectory on identical hardware. This does not make a language model perfectly deterministic, but it removes the largest source of run to run variation, which is what the causal argument needs.

\paragraph{Documents shown to the agent.}
Throughout the paper, the documents we analyze are the ones the agent actually read. We take these from the \texttt{shown\_doc\_ids} field of the trajectory log rather than recomputing the top 10 from the ranked list. The two differ because \texttt{shown\_doc\_ids} is deduplicated across steps. Once a document has been shown at step $t$ it is not shown again later. The reasoning behind that design is that an agent which failed to extract what it needed on first reading should not get a second free look at the same text. A document that the agent never saw cannot have influenced it, so unshown documents are outside the scope of any causal claim we make.

\subsection{Question sampling}
\label{sec:method-sampling}

We draw a reproducible stratified sample of 1000 questions from the 7{,}405 question HotpotQA development set~\citep{yang2018hotpotqa} using the following procedure.

\paragraph{Coverage filter.}
We keep a question only when every article title in its \texttt{supporting\_facts} field appears, matched case insensitively, among the top 50 BM25 results for that question. Questions failing this test are discarded, because the evidence needed to answer them is not reachable within the candidate set at all. Studying agent behavior on questions whose evidence the retriever cannot surface would confound a retrieval failure with a reasoning failure.

\paragraph{Stratified allocation.}
Let $B$ and $C$ be the number of surviving bridge and comparison questions and let $N = B + C$. If $N$ does not exceed the target sample size, every filtered question is kept. Otherwise we allocate the sample to hold roughly 70\% bridge and 30\% comparison questions, which approximates the composition of the development set itself (5{,}918 bridge against 1{,}487 comparison). If one pool cannot fill its quota, we take all of it and top up from the other pool until the target is reached. A fixed random seed makes the draw reproducible. The procedure emits the sampled questions with all original fields preserved, together with a report broken down by question type and difficulty level.

\subsection{Counterfactual trajectory exploration}
\label{sec:method-cte}

Counterfactual Trajectory Exploration (CTE) is our framework for asking what would have happened if an intermediate decision had gone differently. At any reasoning or retrieval step, CTE intervenes on a prior decision and estimates how that change would have propagated through the rest of the trajectory. The aim is to quantify the causal contribution of intermediate decisions to downstream reasoning quality, retrieval efficiency, factual correctness, and task success.

This paper instantiates CTE with a single intervention.

\paragraph{Omission intervention.}
We delete one document from the ranked list at one step and give the agent the remaining $k-1$:
\begin{equation}
  [d_1, d_2, \ldots, d_{10}] \;\longrightarrow\; [d_1, \ldots, d_{i-1}, d_{i+1}, \ldots, d_{10}].
  \label{eq:omission}
\end{equation}
The trajectory is then replayed from that step onward, reusing the logged prefix so that everything before the intervention is identical by construction. The question it answers is deliberately narrow. What if this document had simply not been in the candidate set? Whatever the agent does differently from that point on is attributable to the deleted document and to nothing else. We sweep $i$ over all ten shown documents and $t$ over all search steps, which is what produces a document level dataset rather than a handful of anecdotes.

\subsection{Document level signals}
\label{sec:method-signals}

For every document retrieved at every step of every trajectory in the intervention set we compute four signals. Together they form the document level dataset behind the quadrant analysis.

\subsubsection{Static RAG Utility}
\label{sec:method-sru}

Static RAG Utility (SRU) formalizes the conventional notion of a useful document. Give the document and the question to a reader with no history, no earlier turns, and no accumulated trajectory, then ask how much the answer improves. We prompt the reader to answer using only the supplied document, to return the answer span alone, and to say \texttt{UNANSWERABLE} when the document does not support an answer. The reader prompt appears in Appendix~\ref{app:prompts}.

For question $q_i$ and each document $d_j$ in the set $D_i$ of documents shown across that question's trajectory, we obtain an answer with the document,
\begin{equation}
  \mathrm{SRU}_{\text{with}}(q_i, d_j) = A_{i,j}^{\text{with}},
\end{equation}
and separately an answer with no document at all, which measures what the reader already knows from its parameters,
\begin{equation}
  \mathrm{SRU}_{\text{without}}(q_i) = A_{i}^{\text{without}}.
\end{equation}
The static utility of the document is the improvement it produces over that parametric baseline,
\begin{equation}
  \Delta \mathrm{SRU}(q_i, d_j) = \mathrm{F1}\big(A_{i,j}^{\text{with}}\big) - \mathrm{F1}\big(A_{i}^{\text{without}}\big).
  \label{eq:dsru}
\end{equation}
Subtracting the zero shot baseline matters. Without it, a document would be credited for an answer the reader could have produced from memory.

\subsubsection{Counterfactual Trajectory Utility}
\label{sec:method-ctu}

CTU asks a different question than SRU. Not \emph{does this document help on its own}, but \emph{would the agent have done just as well without it}. We compute it from the omission intervention of Equation~\ref{eq:omission}, scoring the counterfactual against the base trajectory on three axes.

\begin{align}
  \Delta_{\text{answer}} &= \mathrm{F1}_{\text{base}} - \mathrm{F1}_{\text{cf}}, \label{eq:danswer}\\
  \Delta_{\text{next-query}} &= \mathrm{nDCG@10}\big(q_{t+1}^{\text{base}}\big) - \mathrm{nDCG@10}\big(q_{t+1}^{\text{cf}}\big), \label{eq:dnextq}\\
  \Delta_{\text{effort}} &= \mathrm{turns}_{\text{cf}} - \mathrm{turns}_{\text{base}}. \label{eq:deffort}
\end{align}

All three are oriented so that a positive value means the document helped.

$\Delta_{\text{answer}}$ is the most direct signal. Did removing the document make the final answer worse? If so the document was causally important. $\Delta_{\text{next-query}}$ is the local counterpart. Did removing the document push the agent into a different next query that retrieves fewer gold supporting documents? This is the term that makes bridge evidence measurable, because it scores a document on what it caused the agent to search for rather than on what it contained. $\Delta_{\text{effort}}$ asks whether removal cost the agent extra turns, or a successful answer altogether. A document that helps the agent converge quickly shows up here.

\paragraph{Combining the three.}
The components live on different scales, so we min-max normalize each independently across all documents in the intervention set and then sum them with unit weights:
\begin{equation}
  \mathrm{CTU} = w_1 \tilde{\Delta}_{\text{answer}} + w_2 \tilde{\Delta}_{\text{next-query}} + w_3 \tilde{\Delta}_{\text{effort}},
  \qquad w_1 = w_2 = w_3 = 1,
  \label{eq:ctu}
\end{equation}
where $\tilde{\cdot}$ denotes the normalized component. We keep the weights uniform deliberately. Tuning them against the outcome we are trying to demonstrate would make the result circular.

\paragraph{The zero effect point.}
\label{sec:method-null}
Equation~\ref{eq:ctu} has a property worth stating explicitly, because it gives the high and low CTU split a meaning that a median cut alone would not. The observed raw ranges are $[-1, 1]$ for $\Delta_{\text{answer}}$, $[-1, 1]$ for $\Delta_{\text{next-query}}$, and $[-2, 3]$ for $\Delta_{\text{effort}}$. A document that changes nothing at all has a raw value of zero on every component, which normalizes to
\begin{equation}
  \underbrace{\tfrac{0 - (-1)}{1 - (-1)}}_{0.5} + \underbrace{\tfrac{0 - (-1)}{1 - (-1)}}_{0.5} + \underbrace{\tfrac{0 - (-2)}{3 - (-2)}}_{0.4} \;=\; 1.4 .
  \label{eq:null}
\end{equation}
So $\mathrm{CTU} = 1.4$ is exactly the point of no causal effect. Values above it mean the document was, on balance, helping. Values below it mean the agent did better without it. As Section~\ref{sec:results-descriptive} shows, the empirical median of CTU is also $1.400$, which is not a coincidence: the median document has a raw delta of zero on all three components. Thresholding at $1.400$ therefore separates documents that helped from documents that did not, and it happens to coincide with the median rather than being defined by it.

\paragraph{What CTU is a property of.}
Documents do not change trajectories on their own. What changes a trajectory is a model's reading of a document, and two models can reasonably disagree about the same text. The honest form of the quantity is therefore
\begin{equation}
  \mathrm{CTU}(D_t, A, t),
  \label{eq:ctu-agent}
\end{equation}
where $A$ is the reasoning agent and $t$ carries the internal state $(\text{thought} + \text{CW})_t$ at that step. We do not carry $q_t$ separately because it is already encoded in that state. Every number in this paper is measured with respect to one agent, and we make no claim that CTU transfers unchanged to another.

\subsubsection{Retrieval score proxies}
\label{sec:method-proxies}

BM25 scores are unbounded above, so we normalize them by the per question maximum,
\begin{equation}
  \mathrm{BM25}_{\text{norm}}(d_j) = \frac{\mathrm{BM25}(d_j)}{\max_{d \in D_i} \mathrm{BM25}(d)} .
\end{equation}
Cross encoder scores can be negative and their range depends on the model, so we min-max normalize them within each question,
\begin{equation}
  \mathrm{CE}_{\text{norm}}(d_j) = \frac{\mathrm{CE}(d_j) - \min_{d \in D_i} \mathrm{CE}(d)}{\max_{d \in D_i} \mathrm{CE}(d) - \min_{d \in D_i} \mathrm{CE}(d)} .
\end{equation}

\subsubsection{Answer containment}
\label{sec:method-acs}

Answer containment asks the blunt question of whether the document contains the gold answer. We check whether the gold answer string appears in the document text as a case insensitive substring, and whether any gold supporting fact title from the HotpotQA metadata matches the document title. The score is binary rather than graded. It is a strict proxy, included because it is the crudest version of the assumption we are testing.

\subsection{The quadrant construction}
\label{sec:method-quadrant}

The quadrant experiment is the empirical core of the argument. It crosses static utility against causal utility and asks two questions. How often do they disagree, and in which direction?

We split each axis at a threshold and read off four cells. A document is high SRU when $\Delta \mathrm{SRU} > 0$, meaning it improved the stateless reader's answer, and high CTU when $\mathrm{CTU} > 1.400$, meaning its removal hurt the agent on balance, per Equation~\ref{eq:null}.

\begin{table}[t]
  \centering
  \caption{Quadrant cell definitions. The SRU axis splits at zero improvement over the parametric baseline. The CTU axis splits at the zero effect point of Equation~\ref{eq:null}.}
  \label{tab:cell-defs}
  \begin{tabular}{clll}
    \toprule
    Cell & $\Delta \mathrm{SRU}$ & CTU & Interpretation \\
    \midrule
    A & $> 0$ (high) & $> 1.400$ (high) & Expected useful \\
    B & $> 0$ (high) & $\leq 1.400$ (low) & Redundant evidence \\
    C & $\leq 0$ (low) & $> 1.400$ (high) & \textbf{Bridge evidence} \\
    D & $\leq 0$ (low) & $\leq 1.400$ (low) & Genuinely useless \\
    \bottomrule
  \end{tabular}
\end{table}

Table~\ref{tab:cell-defs} names the cells. Two of them are uninteresting by design. Cell A is where both measures agree that a document is useful, which is what everyone already believes good retrieval does. Cell D is where both agree it is noise. The argument lives in the off-diagonal.

\textbf{Cell B, redundant evidence,} is where static utility overestimates. The document contains the answer and the stateless reader is happy, but the agent was going to succeed regardless, either because it had picked the fact up at an earlier step or because the information was already in the model's parameters. Take the Hawaii question again. At step 1 the agent reads that Lingle became the first Republican Governor of Hawaii since 1962, so it already knows the party. If at step 3 it retrieves another passage about Lingle's affiliation, a fresh reader scores that passage highly, yet omitting it changes nothing. The passage is real evidence and causally inert at the same time.

\textbf{Cell C, bridge evidence,} is the cell that carries the paper. Static utility says the document is useless because it does not contain the answer. CTU says the trajectory falls apart without it. The Linda Lingle passage from Section~\ref{sec:introduction} sits here. For the argument to be defensible this cell has to be populated at a non-trivial rate, and we set a threshold of at least 20\% in advance.

\subsection{Entity propagation}
\label{sec:method-propagation}

The quadrant experiment establishes that bridge documents exist. It does not explain why they work. Our account is that a bridge document hands the agent an entity, and the agent puts that entity into its next query. If that is right, the entities that get carried forward should be the ones that carry a real retrieval signal, not any entity that happens to be on the page.

We test this with the Observable Entity Relevance measure of~\citet{ghosh2026entity}, reusing that work's computation without modification.

\paragraph{Hypotheses.}
\begin{itemize}
  \item \textbf{H1 (propagation gap).} An entity that is observably discriminative in a step's candidate set appears in the agent's next query more often than an entity found only in non-relevant documents. We fix a target ratio above $2.5\times$ in advance so a small gap does not count as support.
  \item \textbf{H2 (bridge link).} Among discriminative entities, those carried by bridge documents propagate more than those carried by grounded Cell A documents, which would localize the redirecting entities to the bridge cell.
\end{itemize}

\subsubsection{Relevance labels}
\label{sec:method-qrels}

OER is defined against relevance labels, so we need judgments for every candidate document. We derive them straight from the HotpotQA supporting facts, and two alternatives are worth explaining away.

We do not use the BEIR or \texttt{ir\_datasets} release of HotpotQA. Those index a different version of the corpus with different document identifiers than our paragraph index, which would produce silent zero overlap between run and qrels. Deriving labels from the supporting facts guarantees alignment with our index by construction.

We also do not use a language model to judge relevance. Our agent is a language model. Scoring relevance with an LLM and then testing whether high scoring entities reappear in that same LLM's next query would mostly measure the model's agreement with itself. That circularity would hollow out the result, so the relevance signal has to come from the dataset.

Each candidate paragraph gets one of three levels. Level 2 means the paragraph comes from a gold supporting article and contains the specific gold sentence. Level 1 means it comes from a gold supporting article but not that sentence. Level 0 is everything else and is omitted from the qrels under the usual TREC convention, with unjudged candidates treated as non-relevant. The high and low split uses any level at or above 1, so the level 2 refinement does not affect the partition.

\subsubsection{Unit of analysis and candidate sets}

We treat each agent search step as one OER query, identified as \texttt{\{qid\}\_step\{k\}}, and key both the qrels and the run by that identifier so they stay consistent. The candidate pool for a step is the top 20 documents by cross encoder rank unioned with the documents actually shown at that step. The union guarantees that every document we later measure for propagation carries an OER score, even when cross step deduplication pushes a shown document past rank 20 of its own step.

\subsubsection{Entity linking and OER}

We link every unique candidate document with the WAT entity linker~\citep{piccinno2014tagme} and keep mentions with confidence $\rho \geq 0.1$. WAT returns underscore joined titles, which we normalize to spaces before matching.

For an entity $e$ in the candidate set of step $q$, let $R$ and $NR$ be the counts of relevant and non-relevant candidates, let $\mathrm{df}_{\mathrm{rel}}$ and $\mathrm{df}_{\mathrm{nonrel}}$ be the number of relevant and non-relevant candidates containing $e$, and let $\mathrm{df}_{\mathrm{cand}}$ be the total number of candidates containing $e$. OER is a support weighted difference of smoothed log odds:
\begin{align}
  \hat{p}_{\mathrm{rel}} &= \frac{\mathrm{df}_{\mathrm{rel}} + \alpha}{R + 2\alpha},
  &\hat{p}_{\mathrm{nonrel}} &= \frac{\mathrm{df}_{\mathrm{nonrel}} + \alpha}{NR + 2\alpha}, \label{eq:oer-rates}\\
  w(e,q) &= 1 - \exp\!\left(-\frac{\mathrm{df}_{\mathrm{cand}}}{\tau}\right),
  &\mathrm{OER}(e,q) &= w(e,q)\big[\operatorname{logit}\hat{p}_{\mathrm{rel}} - \operatorname{logit}\hat{p}_{\mathrm{nonrel}}\big]. \label{eq:oer}
\end{align}
The weight $w$ discounts entities seen in only one or two documents, where the log odds estimate would be unstable. We use the source paper's defaults, $\alpha = 0.5$ and $\tau = 5.0$. A positive OER means the entity concentrates in the relevant documents of that step.

We form two contrasting groups. \textbf{High OER} entities appear in at least one gold supporting document ($\mathrm{df}_{\mathrm{rel}} \geq 1$) and are positively discriminative ($\mathrm{OER} > 0$). \textbf{Low OER} entities never appear in a gold supporting document ($\mathrm{df}_{\mathrm{rel}} = 0$). Entities present in gold documents but not discriminative form a middle group that we record and exclude from the contrast. We use this rule rather than a median split because the number of relevant documents per step is small, which makes a median cut unstable.

\subsubsection{Measuring propagation}

Propagation is measured only over documents the agent actually saw, and only into the immediately following query. An entity in a retrieved but unshown document cannot influence the model. The last search step of each trajectory is skipped because it has no next query. We normalize entity titles and queries by lower casing, turning underscores into spaces, and dropping non alphanumeric characters, then apply two criteria.

\textbf{Exact} match, the primary and conservative one, requires the full normalized title to appear as a whole phrase in the next query, matched at token boundaries so that a short title such as \emph{law} cannot match inside \emph{lawyer}. \textbf{Partial} match requires any non-stopword token of length at least two from the title to appear as a whole token, using a 132 word English stoplist. Partial raises the absolute rate and admits common token noise such as \emph{lake} or \emph{county}, which shrinks the gap between groups.

For each criterion we report the propagation rate per group, the ratio between groups, a bootstrap 95\% confidence interval from 2{,}000 resamples, and a chi square test of independence. We treat H1 as supported when the ratio clears $2.5\times$, the test is significant, and the intervals do not overlap.

%% file: sections/results.tex
\section{Results and Analysis}
\label{sec:results}

\subsection{Dataset and descriptive statistics}
\label{sec:results-descriptive}

Before analyzing the intervention set we drop one group. Trajectories that finished in a single turn ($n = 74$ trajectories) have no subsequent step for a document to causally affect, which makes CTU undefined rather than merely small. Keeping them would pad the denominator with cases the measure cannot speak to, so they are removed. The remaining multi-step trajectories yield 23{,}322 document observations covering 16{,}841 unique question and document pairs.

Table~\ref{tab:descriptive} reports the distribution of every signal over the full set. The composite CTU of Equation~\ref{eq:ctu} spans $[0.000, 2.800]$ against a theoretical range of $[0, 3]$.

\begin{table*}[t]
  \centering
  \caption{Descriptive statistics over all 23{,}322 document observations. Every component delta has a median of exactly zero, which is what places the median CTU at the zero effect point of Equation~\ref{eq:null}.}
  \label{tab:descriptive}
  \begin{tabular}{lrrrrr}
    \toprule
    Metric & Mean & Median & Std & Min & Max \\
    \midrule
    $\Delta \mathrm{SRU}$ (\texttt{sru\_f1\_doc\_delta}) & 0.0048 & 0.0000 & 0.1485 & $-1.0000$ & 1.0000 \\
    \texttt{sru\_f1\_withoutdocs} & 0.0122 & 0.0000 & 0.0938 & 0.0000 & 1.0000 \\
    \texttt{sru\_f1\_withdocs} & 0.0170 & 0.0000 & 0.1152 & 0.0000 & 1.0000 \\
    CTU (composite) & 1.3911 & 1.4000 & 0.3137 & 0.0000 & 2.8000 \\
    $\Delta_{\text{answer}}$ & $-0.0225$ & 0.0000 & 0.4117 & $-1.0000$ & 1.0000 \\
    $\Delta_{\text{next-query}}$ & $-0.0482$ & 0.0000 & 0.4427 & $-1.0000$ & 1.0000 \\
    $\Delta_{\text{effort}}$ & 0.1324 & 0.0000 & 1.0631 & $-2.0000$ & 3.0000 \\
    \texttt{bm25\_normalized} & 0.7953 & 0.8091 & 0.1454 & 0.1775 & 1.0000 \\
    \texttt{ce\_normalized} & 0.3700 & 0.2733 & 0.3324 & 0.0000 & 1.0000 \\
    \bottomrule
  \end{tabular}
\end{table*}

Two features of Table~\ref{tab:descriptive} deserve comment before we go further, because both shape how the quadrant should be read.

First, the median of all three CTU components is exactly zero. The typical document changes the answer not at all, changes the next query not at all, and changes the turn count not at all. That is what puts the empirical median of CTU at $1.4000$, matching the analytic zero effect point derived in Equation~\ref{eq:null} to four decimal places. The threshold we use is therefore not a convenience. It is the value at which a document stops mattering.

Second, the static utility axis is compressed. The mean $\Delta \mathrm{SRU}$ is $0.0048$ and its median is zero, and the reader's absolute answer quality is low with or without the document ($0.0170$ against $0.0122$). Answer containment tells the same story from another angle: only 6.38\% of unique pairs and 7.50\% of records contain the gold answer at all. This is expected for multi-hop questions, where most individual paragraphs hold one hop of a two hop chain and cannot support the answer alone. It nevertheless means the SRU axis has little dynamic range, and Section~\ref{sec:results-threats} takes that seriously.

\subsection{The quadrant: static utility against causal utility}
\label{sec:results-quadrant}

Table~\ref{tab:quadrant} crosses the two axes. The headline is the bridge cell. Across all observations, 35.72\% of the documents the agent read were causally load bearing while offering nothing to a stateless reader. That clears the 20\% threshold we committed to in Section~\ref{sec:method-quadrant} by a wide margin. The two cells where the measures agree account for a combined 62.10\%, and the redundant cell, where static utility overestimates, holds another 2.18\%.

\begin{table*}[t]
  \centering
  \caption{Quadrant cell counts and within group percentages. The turn count columns group by the length of the \emph{counterfactual} trajectory, a variable that is confounded with CTU by construction (Section~\ref{sec:results-threats}). They are reported for completeness and no trend should be read from them.}
  \label{tab:quadrant}
  \begin{tabular}{llrrrr}
    \toprule
    Cell & Meaning & All & 2 turns & 3 turns & 4 turns \\
    \midrule
    A & Expected useful   & 262 (1.12\%)    & 50 (0.75\%)    & 75 (1.37\%)    & 137 (1.23\%) \\
    B & Redundant         & 508 (2.18\%)    & 179 (2.67\%)   & 91 (1.66\%)    & 238 (2.14\%) \\
    C & \textbf{Bridge}   & \textbf{8{,}331 (35.72\%)} & 1{,}496 (22.30\%) & 2{,}206 (40.18\%) & 4{,}629 (41.62\%) \\
    D & Genuinely useless & 14{,}221 (60.98\%) & 4{,}985 (74.29\%) & 3{,}118 (56.79\%) & 6{,}118 (55.01\%) \\
    \midrule
    \multicolumn{2}{l}{\textbf{Total}} & \textbf{23{,}322} & \textbf{6{,}710} & \textbf{5{,}490} & \textbf{11{,}122} \\
    \bottomrule
  \end{tabular}
\end{table*}

Table~\ref{tab:quadrant} also breaks the cells out by turn count, and that breakdown needs a warning rather than an interpretation. The bridge cell does rise across the columns, from 22.30\% at two turns to 40.18\% at three and 41.62\% at four, and it is tempting to read this as evidence that harder questions lean more heavily on bridge evidence. We looked into that reading and it does not hold. The grouping variable is confounded with CTU by construction, for reasons Section~\ref{sec:results-threats} sets out in full, and the confound is large enough to account for the entire trend. We report the columns for transparency and draw no conclusion from them. Nothing else in this paper depends on the breakdown.

\subsection{Static and causal utility are close to independent}
\label{sec:results-independence}

The cell counts show disagreement. The rank correlation shows how complete the disagreement is. Table~\ref{tab:spearman} reports Spearman $\rho$ between $\Delta \mathrm{SRU}$ and CTU.

\begin{table}[t]
  \centering
  \caption{Spearman correlation between static utility and counterfactual utility. The coefficient sits near zero in every group. Large $n$ makes the pooled $p$ value small while the effect size stays negligible.}
  \label{tab:spearman}
  \begin{tabular}{lrrr}
    \toprule
    Group & $\rho$ & $p$ & $n$ \\
    \midrule
    All records & $-0.0257$ & $8.754 \times 10^{-5}$ & 23{,}322 \\
    2 turns     & $-0.0083$ & $0.499$                & 6{,}710 \\
    3 turns     & $-0.0252$ & $0.0622$               & 5{,}490 \\
    4 turns     & $-0.0301$ & $0.0015$               & 11{,}122 \\
    \bottomrule
  \end{tabular}
\end{table}

Every coefficient sits within $0.031$ of zero. The pooled correlation reaches statistical significance only because 23{,}322 observations will detect almost any deviation from zero, and a $\rho$ of $-0.026$ explains roughly 0.07\% of the variance. The right reading is not that static utility is mildly negatively related to causal utility. It is that the two are unrelated. A retrieval system that perfectly optimizes static utility carries almost no information about which documents its agent actually needed.

This is the paper's central result, and it is worth being precise about which part is doing the work. The independence is the finding. The size of the bridge cell follows from that independence together with how the two marginals happen to fall, a point we return to in Section~\ref{sec:results-threats}.

\subsection{Robustness: the BM25 and cross encoder quadrant}
\label{sec:results-proxy}

A reader based axis is one way to operationalize static utility. It is not the only one, and it is the one with the least dynamic range in our data. We therefore repeat the quadrant with the static axis replaced by the retrieval scores themselves, which is closer to what a deployed system optimizes and gives an axis that is spread out by construction.

We build a combined proxy from the normalized BM25 and cross encoder scores using global cutoffs (BM25 75th percentile $0.9070$ and median $0.8091$, cross encoder 75th percentile $0.6061$ and median $0.2733$). A document counts as proxy high when both scores agree that it is high and proxy low when both agree it is low. Records where the two signals disagree are excluded, which removes 13{,}722 of 23{,}322 observations (58.84\%) and leaves 9{,}600. The exclusion is substantial and we flag it as a limitation in Section~\ref{sec:results-threats}, but the surviving records give a cleaner test of whether the effect depends on the reader.

\begin{table*}[t]
  \centering
  \caption{Proxy quadrant using normalized BM25 and cross encoder scores as the static axis, over the 9{,}600 records where the two signals agree. The bridge cell remains large. The turn count columns carry the same confound as Table~\ref{tab:quadrant} and are not interpreted.}
  \label{tab:proxy-quadrant}
  \begin{tabular}{llrrrr}
    \toprule
    Cell & Meaning & All & 2 turns & 3 turns & 4 turns \\
    \midrule
    A & Proxy high, CTU high & 957 (9.97\%)   & 153 (5.49\%)  & 253 (11.27\%) & 551 (12.06\%) \\
    B & Proxy high, CTU low  & 1{,}648 (17.17\%) & 570 (20.45\%) & 357 (15.91\%) & 721 (15.78\%) \\
    C & \textbf{Bridge}      & \textbf{2{,}607 (27.16\%)} & 458 (16.43\%) & 690 (30.75\%) & 1{,}459 (31.93\%) \\
    D & Proxy low, CTU low   & 4{,}388 (45.71\%) & 1{,}606 (57.62\%) & 944 (42.07\%) & 1{,}838 (40.23\%) \\
    \midrule
    \multicolumn{2}{l}{\textbf{Total}} & \textbf{9{,}600} & \textbf{2{,}787} & \textbf{2{,}244} & \textbf{4{,}569} \\
    \bottomrule
  \end{tabular}
\end{table*}

The picture holds. Table~\ref{tab:proxy-quadrant} puts the bridge cell at 27.16\%, still above our 20\% bar, on an axis where 27.14\% of documents score high rather than 3.30\%. Rank correlations between the proxy and CTU again sit near zero, and this time most of them do not even reach significance. Over all records $\rho = -0.0161$ with $p = 0.116$. Broken out by turn count the coefficients are $-0.0592$ at two turns, $-0.0349$ at three, and $0.0004$ at four, of which only the two turn figure is significant ($p = 0.0018$).

This matters more than the raw number. The reader based quadrant could be dismissed as an artifact of a weak reader. The proxy quadrant cannot, because it never consults a reader. Whatever a BM25 and cross encoder stack prefers is, on this evidence, unrelated to what the agent needed.

\subsection{Entity propagation: why bridge documents work}
\label{sec:results-propagation}

The quadrant establishes that bridge documents are common. The propagation experiment tests our explanation of them.

The pipeline produced 227{,}139 entity at a step observations from 769 multi-step trajectories, of which 68{,}546 fall in the high OER group, 145{,}927 in the low OER group, and 12{,}666 in the excluded middle. That is close to four times the 60{,}053 observations of an earlier 300 question pilot, which lets us treat the pilot as a first run and these numbers as its replication.

\paragraph{H1 holds, and the ratio is stable across samples.}
Table~\ref{tab:h1} gives the result. Under the conservative exact criterion, an entity that discriminates relevant from non-relevant candidates reaches the agent's next query 6.1\% of the time, against 1.5\% for an entity found only in non-relevant documents. That is a ratio of $4.02\times$, comfortably past the $2.5\times$ bar set in advance. Under the looser partial criterion the rates rise to 14.4\% and 8.2\% for a ratio of $1.77\times$. Both gaps carry a chi square $p$ below $10^{-300}$ with non-overlapping bootstrap intervals.

\begin{table*}[t]
  \centering
  \caption{H1 propagation gap on the 1000 question sample, with the 300 question pilot for comparison. High OER entities reach the agent's next query several times more often than low OER entities, and the ratio barely moves between samples.}
  \label{tab:h1}
  \begin{tabular}{llcccc}
    \toprule
    Criterion & Sample & High OER & Low OER & Ratio & $\chi^2$ $p$ \\
    \midrule
    Exact (whole title)   & 300q pilot     & 8.9\%  & 2.1\%  & $4.18\times$ & $1.6 \times 10^{-271}$ \\
    Exact (whole title)   & \textbf{1000q} & \textbf{6.1\%} & \textbf{1.5\%} & $\mathbf{4.02\times}$ & $< 10^{-300}$ \\
    Partial (title token) & 300q pilot     & 17.9\% & 11.1\% & $1.61\times$ & $7.7 \times 10^{-107}$ \\
    Partial (title token) & \textbf{1000q} & \textbf{14.4\%} & \textbf{8.2\%} & $\mathbf{1.77\times}$ & $< 10^{-300}$ \\
    \bottomrule
  \end{tabular}
\end{table*}

The ratio moves from $4.18\times$ to $4.02\times$ under exact match and from $1.61\times$ to $1.77\times$ under partial match as the sample triples. Absolute rates fall somewhat on the larger sample, which tracks a smaller relevance pool per step (2.9 gold paragraphs on average against 6.1 in the pilot), leaving a high OER entity fewer gold documents to earn its status from. The separation between groups is what stays fixed, and the separation is the claim.

The two criteria trade off exactly as expected. Exact match is a conservative floor that demands the agent reproduce a full entity title verbatim. Partial match allows paraphrase and picks up more real propagation, at the cost of admitting common tokens such as \emph{lake} or \emph{county} that would appear regardless of any OER signal. That noise inflates the low OER rate and compresses the ratio. Exact is the cleaner discrimination signal, partial is the better estimate of the absolute level, and neither reading changes the direction.

Read against the quadrant, this closes the loop. A bridge document is causally essential because it supplies an entity, and the entities that reach the next query are the discriminative ones by a factor of four. The same log odds signal that separates useful entities from generic bait in static reranking~\citep{ghosh2026entity} also governs what a search agent carries forward. The link between static entity relevance and agent behavior is mechanical, not rhetorical.

\paragraph{H2 does not replicate, and we do not claim it.}
Our second hypothesis, that the propagating entities sit specifically inside bridge documents, does not survive at scale. In the pilot, high OER entities in bridge documents propagated slightly more than those in grounded Cell A documents under exact match (9.6\% against 7.9\%). On the full sample the comparison reverses to 5.1\% against 6.2\%, and Cell A holds only 321 high OER observations, which leaves it badly underpowered. Under partial match the bridge cell edges ahead again (11.9\% against 11.2\%). A result that changes sign with the matching rule and rests on a few hundred observations is not one we are willing to build on, so we report H2 as inconclusive. One further caveat sharpens that verdict rather than softening it. The cell assignment used here thresholds CTU at $1.600$ instead of the $1.400$ zero effect point defended in Section~\ref{sec:method-null}, so this partition is stricter than the one behind Table~\ref{tab:quadrant} and the two are not strictly comparable. Since we are declining to claim H2 either way, the discrepancy does not affect any conclusion we draw, but it should be reconciled before anyone revisits the hypothesis.

\begin{table}[t]
  \centering
  \caption{Propagation rate by quadrant cell over all entities. The bridge cell does not stand out, because bridge documents are entity rich and dominated by low OER entities that dilute the average. Two caveats apply. These cells are cut at $\mathrm{CTU} > 1.600$ rather than the $1.400$ used everywhere else in the paper, so they are not directly comparable to Table~\ref{tab:quadrant}. The four counts sum to 223{,}024, with the remaining 4{,}115 observations sitting on documents that carry no cell assignment.}
  \label{tab:cell-propagation}
  \begin{tabular}{llrcc}
    \toprule
    Cell & Meaning & $n$ & Exact rate & Partial rate \\
    \midrule
    A & Expected useful   & 737      & 0.058 & 0.114 \\
    B & Redundant         & 6{,}118  & 0.064 & 0.166 \\
    C & \textbf{Bridge}   & 38{,}860 & 0.032 & 0.093 \\
    D & Genuinely useless & 177{,}309 & 0.045 & 0.122 \\
    \bottomrule
  \end{tabular}
\end{table}

Table~\ref{tab:cell-propagation} shows why the cell cut fails where the OER cut succeeds. Bridge documents are entity rich, and most of the entities on them are low OER. Averaging propagation over every entity in a bridge document therefore dilutes the contribution of the few discriminative ones until the cell looks unremarkable. The informative partition is by OER, not by cell, and that partition is exactly where the effect is clean. The honest summary is that OER predicts propagation, while the localization of propagating entities to a particular quadrant cell is not established by this experiment.

\subsection{Threats to validity}
\label{sec:results-threats}

\paragraph{The turn count breakdown is confounded and carries no difficulty interpretation.}
The columns in Tables~\ref{tab:quadrant} and~\ref{tab:proxy-quadrant} group by the number of turns taken by the \emph{counterfactual} trajectory rather than the base trajectory, so they do not measure how hard a question is to begin with. The deeper problem is that this variable is mechanically tied to a component of the quantity it is crossed against. Since $\Delta_{\text{effort}} = \mathrm{turns}_{\text{cf}} - \mathrm{turns}_{\text{base}}$ and the base trajectory runs between one and four turns, fixing the counterfactual turn count at $c$ confines $\Delta_{\text{effort}}$ to the interval $[c-4,\, c-1]$. The observed per group ranges match that prediction exactly, at $[-2, 1]$ for two turns, $[-1, 2]$ for three, and $[0, 3]$ for four. Conditioning on the group therefore drags $\Delta_{\text{effort}}$ upward as the group index rises, which drags CTU upward, which moves documents across the CTU threshold and inflates the bridge cell without anything substantive happening.

The artifact is more than large enough to produce the whole pattern. Mean $\Delta_{\text{effort}}$ climbs from $-0.5514$ at two turns to $0.6040$ at four, a shift of $0.2311$ once normalized, against a total mean CTU shift of only $0.1594$ across the same groups. The confounded component alone over-explains the trend, which means the other two components drift slightly the other way. We therefore withdraw the difficulty reading entirely and report the columns only so that a reader who computes them is not left wondering why we stayed silent. The aggregate results are untouched, because the independence tests, the bridge cell percentage, and the entire propagation experiment are computed without conditioning on this variable.

\paragraph{The static utility axis is skewed, and the bridge percentage inherits that skew.}
This is the objection we would raise first as reviewers, so we raise it ourselves. Only 3.30\% of observations have $\Delta \mathrm{SRU} > 0$, which means the high and low split on the static axis is close to degenerate. Meanwhile 36.85\% of observations have high CTU. If the two axes were exactly independent, the expected size of the bridge cell would be $(1 - 0.0330) \times 0.3685 = 35.63\%$, against an observed 35.72\%. The bridge cell is, to within a tenth of a percentage point, what independence plus these marginals predicts.

We draw two conclusions from that arithmetic rather than one. The first is a caution. The figure of 35.72\% should not be read as a second, separate discovery on top of the independence result. It is largely a restatement of it, and a paper that presented the two as independent confirmations would be double counting. The second is that the underlying claim is undamaged, because independence is precisely the thing we set out to test and the thing the correlations in Table~\ref{tab:spearman} establish directly. The proxy quadrant of Section~\ref{sec:results-proxy} is the stronger evidence on this point, since it reproduces both a large bridge cell and near zero correlation on an axis where 27.14\% of documents score high rather than 3.30\%.

\paragraph{The static reader is weak in absolute terms.}
Answer F1 for the reader averages $0.0170$ with a document and $0.0122$ without one. Most shown paragraphs simply cannot support a standalone answer to a multi-hop question, which is what the 7.50\% answer containment rate reflects. The consequence is that low SRU often means the reader failed both with and without the document, rather than that the document was demonstrably unhelpful. This weakens the reader based axis as a measurement instrument. It does not weaken the argument, because the claim under test is about what the agent needed rather than about what the reader could do, and the proxy quadrant reaches the same conclusion without a reader.

\paragraph{The proxy quadrant discards disagreement.}
Requiring BM25 and the cross encoder to agree excludes 58.84\% of records. Those excluded cases are exactly the ones where the two retrieval signals conflict, and they may well behave differently. The proxy quadrant should be read as a check on the subset where a deployed stack would be confident, not as a claim about the full population.

\paragraph{CTU is a property of an agent, not of a document.}
Equation~\ref{eq:ctu-agent} makes this explicit. Every number here is measured with respect to one backbone, Qwen2.5-7B-Instruct, under one prompt and one retrieval stack. A different model could read the same passage differently and assign it a different CTU. We report a property of an agent and its evidence jointly, and whether the bridge phenomenon holds at other scales is an open question we do not answer.

\paragraph{Determinism is approximate.}
The causal argument assumes the agent responds consistently to the same context. We set temperature to $0.0$, disable sampling, and hold the hardware fixed, which removes the dominant sources of variation but does not make a language model exactly reproducible. Residual non-determinism adds noise to individual CTU scores. Given effect sizes measured over tens of thousands of observations, we do not expect it to move the aggregate conclusions.

\paragraph{The relevance pool for OER is small.}
Our OER estimates come from HotpotQA gold supporting facts, which supply roughly two to six relevant paragraphs per step. That is a smaller pool than the setting the original OER study used, so individual scores are noisier here. The effect nonetheless comes through under both matching criteria and across a threefold change in sample size, which is why we are comfortable resting the cross experiment claim on it.

%% file: sections/conclusion.tex
\section{Conclusion}
\label{sec:conclusion}

We set out to test an assumption that retrieval research rarely states out loud: that a document useful to a reader in isolation is the same document an agent needs. Across 23{,}322 counterfactual document observations drawn from 1000 HotpotQA questions, it is not. Static utility and causal utility correlate at $\rho = -0.026$, which is another way of saying they are unrelated. About a third of the documents our agent read were load bearing while looking worthless to a stateless reader. Swapping the reader for a BM25 and cross encoder proxy, which is closer to what deployed systems actually optimize, leaves the picture intact at 27.16\%.

We also know why these documents work. They carry entities the agent picks up and searches with. Entities that discriminate relevant from non-relevant candidates reach the next query 4.02 times more often than entities found only in non-relevant documents, over 227{,}139 observations. The signal that identifies useful entities for a static ranker~\citep{ghosh2026entity} turns out to identify the entities a search agent propagates. That makes the connection between the two settings mechanical rather than a matter of framing, and it gives bridge evidence a concrete definition instead of an intuition.

Three results did not go our way and we report them as such. The finer hypothesis that propagating entities localize to bridge cells reversed direction between the pilot and the full sample on an underpowered comparison, so we treat it as inconclusive. The reader based static axis turned out to be badly skewed, with only 3.30\% of documents scoring above zero, which means the headline bridge percentage largely restates the independence result rather than confirming it separately. And an appealing secondary pattern, in which the bridge cell grew with the number of turns, dissolved once we checked it. The grouping variable is the counterfactual turn count, which is mechanically tied to the effort term inside CTU, and the induced artifact more than accounts for the trend. We withdrew that claim rather than ship it. What survives is the independence itself, the proxy quadrant, and the propagation result, each of which stands without the others.

What follows from this is a target problem rather than a finished solution. If causal utility is what matters in agentic retrieval and static utility does not predict it, then the labels the field trains on are the wrong labels for this setting, and the metrics it reports are measuring something adjacent to what it wants. The obvious next step is a selector that approximates CTU without running the intervention, since counterfactual replay is far too expensive to sit inside a training loop. The propagation result suggests where to look. Observable entity relevance is computable from candidate set statistics alone, needs no counterfactual, and predicts the behavior we care about by a factor of four. That looks like a usable proxy for the signal a ranker would need in order to serve an agent rather than a reader.

%% file: sections/appendix.tex
\appendix

\section{Prompts}
\label{app:prompts}

All three prompts are reproduced verbatim as used in the experiments.

\subsection{Agent system prompt}
\label{app:agent-prompt}

The ReAct style research agent of Section~\ref{sec:method-trajectory} runs under the following system prompt.

\begin{quote}\small\ttfamily
You are a research agent that answers questions by searching Wikipedia.\\
Break complex questions into steps and search iteratively.\\
Rules:\\
1. Use <thought> to reason before each action.\\
2. Search with <action>search</action> and <query>.\\
3. Answer with <action>answer</action> when you have enough evidence.\\
4. <answer> must be concise: a name, date, place, or short phrase.\\
5. Cite document numbers: <citations>Doc 1, Doc 3</citations>. If none: <citations>NONE</citations>.\\
6. Maximum 4 search turns. If you reach 4, answer with your best guess.\\
7. Never repeat the same query.
\end{quote}

\subsection{Static reader prompt}
\label{app:reader-prompt}

The reader used to compute $\mathrm{SRU}_{\text{with}}$ in Section~\ref{sec:method-sru} is constrained to the supplied document and to a bare answer span.

\begin{quote}\small\ttfamily
Answer using ONLY the information in the provided document. Respond with ONLY the answer itself: a name, date, place, or short phrase (a few words at most). Do NOT explain your reasoning, do NOT restate the question, and do NOT include any text other than the answer span itself. If the document does not contain enough information to answer, you MUST write UNANSWERABLE. Do not use any outside knowledge.\\[4pt]
Example:\\
Question: What year was the Eiffel Tower completed?\\
Document: The Eiffel Tower is a wrought-iron lattice tower in Paris, France. It was completed in 1889 as the entrance arch for the World's Fair.\\
Answer: 1889\\[4pt]
Question: \{question\}\\
Document: \{passage\}\\
Answer:
\end{quote}

\subsection{Zero shot prompt}
\label{app:zeroshot-prompt}

The parametric baseline $\mathrm{SRU}_{\text{without}}$ uses the same reader with no document supplied.

\begin{quote}\small\ttfamily
Answer the following question. Respond with ONLY the answer itself: a name, date, place, or short phrase (a few words at most). Do NOT explain your reasoning, do NOT restate the question, and do NOT include any text other than the answer span itself.\\[4pt]
Example:\\
Question: What year was the Eiffel Tower completed?\\
Answer: 1889\\[4pt]
Question: \{question\}\\
Answer:
\end{quote}